\documentclass[11pt]{article}
\begin{document}
\title{
Two New Integrable Lattice Hierarchies Associated With A Discrete Schr\"{o}dinger Nonisospectral Problem and Their
Infinitely Many Conservation Laws}
\author{
Zuo-nong Zhu $^{1}$ and Weimin Xue $^{2}$\\
1. Department of Mathematics, Shanghai Jiao Tong University,\\
Shanghai, 200030, P.R. China, znzhu@online.sh.cn\\
2. Department of Mathematics, Hong Kong Baptist University,\\
Hong Kong, P. R. China, wmxue@math.hkbu.edu.hk\\
}
\date{}
\maketitle
\begin{abstract}
In this article, by means of using discrete zero curvature 
representation and constructing opportune time evolution problems,
two new discrete integrable lattice hierarchies with n-dependent coefficients are proposed, which related to a new discrete Schr\"{o}dinger nonisospectral operator equation. The relation of the two new lattice hierarchies with the Volterra hierarchy is discussed. It has been shown that one lattice hierarchy is equivalent to the positive Volterra hierarchy with n-dependent coefficients and another lattice hierarchy with isospectral problem is equivalent to the negative Volterra hierarchy.
We demonstrate the existence of infinitely many conservation laws for the two lattice hierarchies 
and give the corresponding conserved densities and the associated fluxes
formulaically.
Thus their integrability is confirmed.
\end{abstract}
\setcounter{section}{0}
\setcounter{equation}{0}
\section{Introduction}
In recent years there has been wide interests in the study of nonlinear integrable lattice systems. It is well known that discrete lattice systems not only have 
rich mathematical structures but also have many applications in science, such as mathematical physics, numerical analysis, statistical physics, quantum physics, etc. Recently, Boiti and co-authors [1] proposed a whole class of nonlinear lattice evolution equations, by use of the Lax technique introduced in [2] and [3], which correspond to isospectral deformations of the new Schr\"{o}dinger discrete spectral operator, 
\begin{eqnarray}
(E^2-q_{n+1}E)
\bar{\psi}_n(\lambda)=\lambda \bar{\psi}_n(\lambda),
\end{eqnarray}
i.e.
\begin{eqnarray}
E\psi_{n}(\lambda)
=U_n(\lambda)\psi_n(\lambda),\qquad U_n(\lambda)=\left(
\begin{array}{cc}
0 & 1 \\
\lambda&q_n
\end{array}  \right),
\end{eqnarray}
where $n\in Z$ is a discrete variable, $\lambda\in C$ is the spectral parameter and $E^k$ is the shift operator defined by $
E^kf(n)=f(n+k), k\in Z$.
This spectral equation was introduced by Shabat in [4] and investigated by Boiti et al in [5]. Integrable lattice hierarchies related to (1.1) are interesting. It has been shown that they contain as special cases discrete versions
of KdV, Sine-Gordon and Liouville equations.  
The Darboux transformation and  the B\"{a}cklund transformation for the proposed lattice hierarchies were also obtained in [1]. In [6], we demonstrated the existence of infinitely many conservation laws for the proposed lattice hierarchies and gave the corresponding conserved densities and the associated fluxes formulaically.\\ 
In this paper, we would like to consider nonisospectral deformations of the Schr\"{o}dinger discrete spectral operator equation (1.1). By means of constructing opportune time evolution equation explicitly,
\begin{equation}
\frac{d\psi_n(\lambda)}{d
t}=V_n^{(m)}(q_n,\lambda)\psi_n(\lambda),
\end{equation}
where $V_n^{(m)}$ is a proper $2\times 2$ matrix, and using the discrete zero curvature representation
\begin{equation}
\frac{\partial U_{n}}{\partial t}+\frac{\partial U_n}{\partial \lambda}\frac{d\lambda}{dt}-(EV_{n}^{(m)})U_n+U_nV_{n}^{(m)}=0,
\end{equation}
where
$\frac{d\lambda}{dt}=a\lambda^{\beta}, a\not=0$
with $\beta$ being a proper constant,
we propose two integrable lattice hierarchies with n-dependent coefficients related to nonisospectral problem (1.2). The relation of the two new lattice hierarchies with the Volterra lattice hierarchy is discussed.
It is well known that the existence of infinitely many conservation laws is very important indicator of integrability of the system. From physical view and numerical analysis, it is also very useful to know whether exist
conservation laws for a lattice system. Infinitely many conservation laws for many discrete lattice systems have been obtained. However, to our knowledge, conservation laws for the lattice system with n-dependent coefficients have not been discussed in the literature.
In this article, using the explicit Lax pairs and following the method studied in [6-10], we will demonstrate the existence of infinitely many conservation laws for the obtained two lattice hierarchies 
and give the corresponding conserved densities and the associated fluxes
formulaically.         
It should be remarked that an extension of the discrete Schr\"{o}dinger spectral problem (1.1), i.e.
\begin{eqnarray}
(E^2+a_nE+b_n+c_nE^{-1})
\psi_n=\lambda \psi_n
\end{eqnarray}
and associated evolution equations were studied in [11,12]. However, the condition $c_n=0$ in above operator equation is not allowed in [11,12].
\setcounter{equation}{0}
\section{Two integrable lattice hierarchies with n-dependent coefficient associated with 
nonisospectral problem (1.2)}
The derivation of new integrable lattice hierarchy is always very important and interesting, though the used method sometimes is standard. In this section, we derive two integrable lattice hierarchies with n-dependent coefficients associated with nonisospectral problem (1.2) by means of discrete zero curvature representation and study the relation of the two lattice hierarchies with the Volterra hierarchy. Let's construct opportune time evolution matrix $V_n^{(m)}$ as follows,
\begin{eqnarray}
V_n^{(m)}=\left(
\begin{array}{cc}
B^{(m)}(\lambda)& A^{(m)}(\lambda)\\
\lambda EA^{(m)}(\lambda)& C^{(m)}(\lambda)
\end{array}\right)
\end{eqnarray}
with
$$A^{(m)}(\lambda)=\sum_{j=-1}^{m}A_j\lambda^{m-j},\qquad B^{(m)}(\lambda)=\sum_{j=-1}^{m}B_j\lambda^{m-j},\qquad C^{(m)}(\lambda)=\sum_{j=-1}^{m}C_j\lambda^{m-j},$$
where $A_j, B_j, C_j (j=-1,0,1.....m)$ are determined by the following equation:
\begin{eqnarray}
(E+1)B_j=-q_nEA_j,\qquad (E^2-1)A_j=q_n(E-1)B_{j-1},\nonumber\\
\qquad C_j=-B_j,\qquad j=0,1,2,...,m\\
 EB_{-1}-C_{-1}+q_nEA_{-1}=0,\qquad (E^2-1)A_{-1}=0,\nonumber\\
 EC_{-1}-B_{-1}-q_nEA_{-1}=a.\nonumber
\end{eqnarray}
Here we suppose time evolution of spectral parameter $\lambda$ is described by $\frac{d\lambda}{dt}=a\lambda^{m+2}, m\geq -1.$
From discrete zero curvature representation, an integrable lattice hierarchy is proposed,
\begin{eqnarray}
\dot{q}_n=q_n(E-1)C_m, \qquad m\geq -1,
\end{eqnarray}
where $C_m, m\geq -1$ can be found from equation (2.2) via the path:
$$A_{-1}\rightarrow B_{-1}\rightarrow C_{-1} \rightarrow A_0\rightarrow C_0\rightarrow........\rightarrow A_{m-1}\rightarrow C_{m-1}\rightarrow A_m\rightarrow C_m\rightarrow.....$$ 
By means of the following formulas:
\begin{eqnarray}
(E+1)^{-1}=\sum_{k=0}^{\infty}(-1)^kE^k,\nonumber\\
(E-1)^{-1}=-\sum_{k=0}^{\infty}E^k,\\
(E^2-1)^{-1}=-\sum_{k=0}^{\infty}E^{2k},\nonumber
\end{eqnarray}
and choose $A_{-1}=-1$, we obtain the solutions to equation (2.2):
\begin{eqnarray*}
B_{-1}=(\frac{n}{2}-\frac{1}{4})a+c_1(-1)^n+\sum_{k=0}^{\infty}(-1)^kq_{n+k},\\
C_{-1}=(\frac{n}{2}+\frac{1}{4})a-c_1(-1)^n-\sum_{k=0}^{\infty}(-1)^{k}q_{n+k},
\end{eqnarray*}
$$A_0
=c_2+c_3(-1)^n+(2c_1(-1)^n-\frac{a}{2})\sum_{k=0}^{\infty}q_{n+2k}+\sum_{k=0}^{\infty}q_{n+2k}^2
-2\sum_{k=0}^{\infty}q_{n+2k}\sum_{j=1}^{\infty}(-1)^{j+1}q_{n+j+2k},$$
\begin{eqnarray}
C_0=c_4(-1)^n+c_2\sum_{k=0}^{\infty}(-1)^kq_{n+k}-c_3(-1)^n\sum_{k=0}^{\infty}q_{n+k}
-\frac{a}{2}\sum_{k=0}^{\infty}(-1)^kq_{n+k}\sum_{j=0}^{\infty}q_{n+2j+k+1}\\
-2c_1(-1)^n\sum_{k=0}^{\infty}q_{n+k}\sum_{j=0}^{\infty}q_{n+2j+k+1}+\sum_{k=0}^{\infty}(-1)^kq_{n+k}\sum_{j=0}^{\infty}q_{n+2j+k+1}^2\nonumber\\
-2\sum_{k=0}^{\infty}(-1)^kq_{n+k}\sum_{j=0}^{\infty}q_{n+2j+1+k}\sum_{i=2}^{\infty}(-1)^{i}q_{n+i+2j+k}\nonumber\\
A_1=(E^2-1)^{-1}[q_n(1-E)C_0],\qquad
C_1=(E+1)^{-1}(q_nEA_1),\nonumber
\end{eqnarray}
$$............$$
where $c_i,(i=1,2,3,4)$ are arbitrary constants.
The first flow and the second flow of lattice hierarchy (2.3) are described, respectively, 
\begin{eqnarray}
\dot{q}_n=q_n(2c_1(-1)^n+\frac{a}{2}+q_n+2\sum_{k=1}^{\infty}(-1)^kq_{n+k}),
\end{eqnarray}
\begin{eqnarray}
\dot{q}_n=q_n(E-1)C_0,
\end{eqnarray}
In order to obtain the second lattice hierarchy, we set, in matrix $V_n^{(m)}$, that 
\begin{eqnarray*}
A^{(m)}(\lambda)=\sum_{j=0}^{m}A_j\lambda^{j-m-1},\qquad B^{(m)}(\lambda)=\sum_{j=0}^{m}B_j\lambda^{j-m-1},\qquad
C^{(m)}(\lambda)=\sum_{j=0}^{m}C_j\lambda^{j-m-1},
\end{eqnarray*}
where
$A_j, B_j, C_j (j=0,1,2,....m)$ are determined by the following equation: 
\begin{eqnarray}
(E+1)B_j=-q_nEA_j,\qquad
(E^2-1)A_{j-1}+q_n(E-1)C_{j}=0,\nonumber\\
B_j=-C_j, \qquad j=1,2,...,m\\
EB_0-C_0+q_nEA_0=0,\qquad (E-1)C_0=0,\nonumber\\
EC_0-B_0-q_nEA_0=a.\nonumber
\end{eqnarray}
Here the time evolution of spectral parameter $\lambda$ is described by
$\frac{d\lambda}{dt}=a\lambda^{-m}, m\geq 0.$ By means of discrete zero curvature representation,
another lattice hierarchy is proposed,
\begin{eqnarray}
\dot{q}_n=(E^2-1)A_m, \qquad m\geq 0,
\end{eqnarray}
where $A_m, m\geq 0$ are determined from equation (2.8) via the path:
$$C_0\rightarrow B_0\rightarrow A_0\rightarrow 
B_1\rightarrow 
A_1\rightarrow.......\rightarrow B_{m-1}\rightarrow A_{m-1}\rightarrow B_m\rightarrow A_m\rightarrow......$$ 
Choosing $C_{0}=0$, we obtain
$$B_0=na-1, \qquad A_{0}=\frac{1-na}{q_{n-1}},\qquad
B_1=\frac{1-na}{q_{n}q_{n-1}}+a\sum_{k=0}^{\infty}\frac{1}{q_{n+k}q_{n+k+1}},$$
\begin{eqnarray}
A_1=\frac{(n-1)a-1}{q_{n-1}^2}(\frac{1}{q_n}+\frac{1}{q_{n-2}})-\frac{2a}{q_{n-1}}\sum_{k=0}^{\infty}\frac{1}{q_{n+k}q_{n+k+1}},
\end{eqnarray}
$$B_2=\frac{-1}{q_{n-1}q_n}[\frac{1+(1-n)a}{q_{n-2}q_{n-1}}+\frac{1+(3-n)a}{q_{n-1}q_{n}}+\frac{1-na}{q_{n}q_{n+1}}]
-\frac{2a}{q_{n-1}q_n}\sum_{k=1}^{\infty}\frac{1}{q_{n+k}q_{n+k+1}}-3a\sum_{k=0}^{\infty}\frac{1}{q_{n+k}^2q_{n+k+1}^2},$$
$$A_2=\frac{-(1+E^{-1})B_2}{q_{n-1}},$$
$$................$$
The first flow of (2.9) is given by  
\begin{eqnarray}
\dot{q}_n=\frac{1-(n+2)a}{q_{n+1}}-\frac{1-na}{q_{n-1}},
\end{eqnarray}
which is just a discrete $KdV$ equation with $n$ dependent coefficient. The second flow of (2.9) is 
\begin{eqnarray}
\dot{q}_n=\frac{(n+1)a-1}{q_{n+1}^2}(\frac{1}{q_{n+2}}+\frac{1}{q_n})-\frac{(n-1)a-1}{q_{n-1}^2}(\frac{1}{q_{n}}+\frac{1}{q_{n-2}})\nonumber\\
-\frac{2a}{q_{n+1}}\sum_{k=2}^{\infty}\frac{1}{q_{n+k}q_{n+k+1}}+\frac{2a}{q_{n-1}}\sum_{k=0}^{\infty}\frac{1}{q_{n+k}q_{n+k+1}}.
\end{eqnarray}
We notice that by considering the following two transformations for (1.1):
\begin{eqnarray}
\bar{\psi}_n=\bar{\phi}_n\lambda^{n/2}\prod_{k=n+1}^{\infty}q_k,\\
\bar{\psi}_n=\bar{\phi}_n\lambda^{n/2}\prod_{k=-\infty}^{n}\frac{1}{q_k},
\end{eqnarray}
the nonisospectral problem (1.1) becomes  
\begin{eqnarray}
(q_{n}q_{n+1})^{-1}\bar{\phi}_{n+1}=\bar{\phi}_{n-1}+\lambda^{-1/2}\bar{\phi}_n
\end{eqnarray}
i.e.,
\begin{eqnarray}
E\phi_n=\bar{U}_n\phi_n, \qquad \bar{U}_n=\left(
\begin{array}{cc}
0&1\\
q_nq_{n+1}&\lambda^{-1/2}q_nq_{n+1}
\end{array}
\right)
\end{eqnarray}
and continuous time evolution equation (1.3) becomes 
\begin{eqnarray}
\frac{d\phi_n(\lambda)}{d
t}=\bar{V}_n^{(m)}(q_n,\lambda)\phi_n(\lambda),
\end{eqnarray}
where
\begin{eqnarray}
\bar{V}_n^{(m)}(q_n,\lambda)=\left(
\begin{array}{cc}
v_{11}^{(m)}-\frac{n-1}{2\lambda}\frac{d\lambda}{dt}-\sum_{k=n}^{\infty}\frac{\dot{q}_k}{q_k}& \frac{\lambda^{1/2}v_{12}^{(m)}}{q_n}\\
\lambda^{-1/2}q_nv_{21}^{(m)}& v_{22}^{(m)}-\frac{n}{2\lambda}\frac{d\lambda}{dt}-\sum_{k=n+1}^{\infty}\frac{\dot{q}_k}{q_k}
\end{array}
\right)
\end{eqnarray}
Discrete spectral problem (2.15) is the well-known Volterra discrete spectral problem with canonical variable $\frac{1}{q_{n}q_{n+1}}$. So, it may be questioned that 
are lattice hierarchies (2.3) and (2.9) equivalent to the Volterra hierarchy with n dependent coefficients? It is well known that lattice hierarchy derived from discrete zero curvature representation 
not only depends on discrete spectral operator equation but also relates to its continuous-time evolution problem. In the following, we will give an answer to the question. By the discrete zero curvature representation of $\bar{U}_n$ and $\bar{V}_n$, we have
\begin{eqnarray}
\frac{d(q_nq_{n+1})}{dt}=q_nq_{n+1}(Ev_{22}^{(m)}-v_{11}^{(m)}-\lambda^{-1}q_nv_{21}^{(m)}-\lambda^{-1}\frac{d\lambda}{dt}+\frac{\dot{q}_n}{q_n}+\frac{\dot{q}_{n+1}}{q_{n+1}}),\\
\lambda^{-1/2}\frac{d(q_nq_{n+1})}{dt}=\lambda^{-1/2}q_nq_{n+1}[(E-1)v_{22}^{(m)}+\frac{1}{q_n}(E-E^{-1})V_{21}^{(m)}+\frac{\dot{q}_{n+1}}{q_{n+1}}]
\end{eqnarray}
For lattice hierarchy (2.9), notice the conditions (2.8) of $v_{ij}^{(m)}$, the equations (2.19) and (2.20) are compatible, which leads to
\begin{eqnarray}
\frac{d(q_nq_{n+1})}{dt}=q_{n+1}(E^2-1)A_m+q_n(E^2-1)EA_m, \qquad m\geq 0.
\end{eqnarray}
However, for lattice hierarchy (2.3), notice the conditions (2.2) of $v_{ij}^{(m)}$,the equations (2.19) and (2.20) are compatible only if $a=0$. In this case, we have
\begin{eqnarray}
\frac{d(q_nq_{n+1})}{dt}=q_nq_{n+1}(E^2-1)C_m, \qquad m\geq -1.
\end{eqnarray}
Now let's discuss the relation of lattice hierarchy (2.21) with the positive Volterra lattice hierarchy. First notice that positive Volterra lattice hierarchy is described by
\begin{eqnarray}
\frac{d u_n}{dt}=u_n(E-1)(1+E^{-1})e_m,\qquad m\geq 0,
\end{eqnarray}
where $e_m$ is determined by the following equation:
\begin{eqnarray}
(E-E^{-1})e_{j}+(E-1)h_{j+1}=0,\qquad j\geq 0\nonumber\\
(E-1)e_j+u_{n+1}E^2h_j-u_nh_j=0, \qquad j\geq 0\\
(E-1)h_0=0,\nonumber
\end{eqnarray}
By introducing $u_n=\frac{1}{q_nq_{n+1}}$, and $t\rightarrow -t$ for even flows, the hierarchy (2.21) is written as
\begin{eqnarray}
\frac{du_n}{dt}=u_n(E-1)(E+1)(-1)^{m+1}C_{m+1}, \qquad  m\geq 0.
\end{eqnarray}
If only considering isospectral problem,
we can prove the following formula by the induction:
\begin{eqnarray}
(-1)^{j+1}(E+1)C_{j+1}=(1+E^{-1})e_j, \qquad j\geq 0.
\end{eqnarray}
In fact, since $e_0=u_n, C_1=-u_{n-1}$, equation (2.26) holds for $j=0$. Suppose it is true
for $j=m-1$, then notice that
\begin{eqnarray}
C_{m+1}=(E-1)^{-1}[q_n^{-1}(1-E^2)(\frac{E^{-1}(1+E)C_m}{q_{n-1}})]\nonumber\\
=(-1)^m(E-1)^{-1}[q_n^{-1}(1-E^2)(\frac{E^{-1}(1+E^{-1})e_{m-1}}{q_{n-1}})]
\end{eqnarray}
and 
\begin{eqnarray}
E^{-1}(1+E^{-1})e_{m-1}=q_{n-1}(E^2-1)^{-1}[q_n(E-1)E^{-1}e_m].
\end{eqnarray}
Then, equation (2.26) is also true for $j=m$. Thus, hierarchy (2.25) is equivalent to positive Volterra hierarchy for isospectral and nonisospectral problems. 
The first and the second nonisospectral flow of the hierarchy (2.25) are described by the following equations, respectively,
\begin{eqnarray}
\frac{d u_n}{dt}=u_n(u_{n+1}-u_{n-1})-au_n[(n+3)u_{n+1}+u_n-nu_{n-1}]
\end{eqnarray}
\begin{eqnarray}
\frac{d u_n}{dt}=u_nu_{n+1}(u_n+u_{n+1}+u_{n+2})-u_nu_{n-1}(u_n+u_{n-1}+u_{n-2})+\nonumber\\
+2au_n^2(u_{n+1}+2\sum_{k=n+2}^{\infty}u_k)+2au_nu_{n+1}\sum_{k=n+3}^{\infty}u_k
+2au_nu_{n-1}\sum_{k=n}^{\infty}u_k\nonumber\\
+nau_n^2(u_n+u_{n-1})
+(n-1)au_{n-1}u_n(u_{n-2}+u_{n-1})\nonumber\\
-(n+1)au_n^2(u_n+u_{n+1})-(n+2)au_nu_{n+1}(u_{n+1}+u_{n+2}).
\end{eqnarray}
Very recently, the negative Volterra hierarchy is proposed by Pritula and Vekslerchik in [13], which has the form,
\begin{eqnarray}
\frac{d u_n}{dt}=u_n(E-1)g_{m+1}, \qquad m\geq -1,
\end{eqnarray}
where $g_i, i\geq 0$ is determined by the following equation:
\begin{eqnarray}
(E-E^{-1})f_{j}+(E-1)g_{j-1}=0,\qquad j\geq 1\nonumber\\
(E-1)f_j+u_{n+1}E^2g_j-u_ng_j=0, \qquad j\geq 1\\
(E-E^{-1})f_0=0,\qquad
u_{n+1}E^2g_0-u_ng_0=0.\nonumber
\end{eqnarray}
Set $u_n=\frac{1}{q_nq_{n+1}}$ in equation (2.32), then $g_0=q_n$. 
Under transformation $u_n=\frac{1}{q_nq_{n+1}}$, and $t\rightarrow -t$ for even flows, the hierarchy (2.22) possesses the form
\begin{eqnarray}
\frac{d u_n}{dt}=u_n(E-1)(E+1)(-1)^mC_m, \qquad m\geq -1.
\end{eqnarray}
The first flow of the hierarchy (2.33) is written as,
\begin{eqnarray}
\frac{d u_n}{dt}=\frac{1}{q_{n}}-\frac{1}{q_{n+1}}
\end{eqnarray}
which is just the simplest flow of negative Volterra hierarchy (2.31). The fact is very interesting. Can we establish relation
between lattice hierarchy (2.33) and the negative Volterra lattice hierarchy (2.31)? Answer is yes.  
In fact, we can prove the following formula by the induction, 
\begin{eqnarray}
(-1)^{j}(E+1)C_j=g_{j+1}, \qquad 
j\geq -1.
\end{eqnarray}
First, from
equation (2.2) we have $-(E+1)C_{-1}=q_n=g_0$, thus equation (2.35) holds as $j=-1$. Notice that 
\begin{eqnarray}
(E+1)C_j=q_n(E^2-1)^{-1}[q_{n+1}(1-E)EC_{j-1}],\nonumber\\
g_{j+1}=q_n(E^2-1)^{-1}[q_{n+1}(1-E)f_{j+1}], \qquad j\geq 0
\end{eqnarray}
So, for $j\geq 0$, equation (2.35) is equivalent to 
\begin{eqnarray}
(-1)^{j} EC_{j-1}=f_{j+1},\qquad j\geq 0
\end{eqnarray}
Since
$$f_1=-(1+E)^{-1}Eg_0=\sum_{k=0}^{\infty}(-1)^{k+1}q_{n+k+1}=EC_{-1},$$
equation (2.37) is true for $j=0$. Suppose equation 
(2.37) holds for $j=m$, then it also holds for $j=m+1$.
In fact, we have
\begin{eqnarray}
f_{m+2}=-(1+E)^{-1}Eg_{m+1}=-(1+E)^{-1}[q_{n+1}(E^2-1)^{-1}(q_{n+2}(1-E)Ef_{m+1})]\nonumber\\
=(-1)^{m+1}(1+E)^{-1}[q_{n+1}(E^2-1)^{-1}(q_{n+2}(1-E)E^2C_{m-1})]
=(-1)^{m+1}EC_{m}
\end{eqnarray} 
From above analysis, we conclude that lattice hierarchy (2.9) is equivalent to the positive Volterra hierarchy with n-dependent coefficients and lattice hierarchy (2.3) with $a=0$ is equivalent to the negative Volterra hierarchy. We thus believe it was worthwhile to study nonisospectral problem (1.2) and the related lattice hierarchies in a independent way.
\setcounter{equation}{0}
\section{Infinitely many conservation laws for lattice hierarchies (2.3) and (2.9)}
For a lattice equation
\begin{eqnarray}
F(\dot{q}_n,\ddot{q}_n,...,q_{n-1},q_n,q_{n+1},...)=0,
\end{eqnarray}
if there exist functions $\rho_n$ and $J_n$, such that
\begin{equation}
\dot{\rho}_n|_{F=0}=J_{n+1}-J_n,
\end{equation}
then equation (3.2) is called the conservation law of equation (3.1),
where $\rho_n$ is
the conserved density and $J_n$ is the associated flux. Suppose equation
(3.1) has conservation law (3.2) and $J_n$ is bounded for all $n$ and vanishes at the boundaries, then $\sum_n\rho_n=c$, with $c$ being arbitrary constant, is an integral of motion of lattice equation (3.1). 
In this section, we first demonstrate the existence of infinitely many conservation laws for lattice hierarchy related to nonisospectral problem (1.2) by means of the explicit Lax pairs, and then we derive infinitely many conservation laws for lattice hierarchies  
(2.3) and (2.9) in details and give the corresponding conserved densities and the associated fluxes formulaically.\\
{\bf 3.1} Infinitely many conservation laws for lattice hierarchy related to nonisospectral problem (1.2)\\
For discrete Schr\"{o}dinger nonisospectral problem (1.2)
\begin{eqnarray}
\psi_{2,n+1}=\lambda\psi_{2,n-1}+q_n\psi_{2,n},
\end{eqnarray}
if set $\Gamma_n=\frac{\psi_{2,n-1}}{\psi_{2,n}}$ and notice that
\begin{eqnarray}
\frac{(\psi_{2,n+1}\psi_{2,n}^{-1})_t}{\psi_{2,n+1}\psi_{2,n}^{-1}}
=\frac{(\psi_{2,n+1})_t}{\psi_{2,n+1}}-\frac{(\psi_{2,n})_t}{\psi_{2,n}},
\end{eqnarray}
then we obtain
\begin{eqnarray}
\frac{\partial}{\partial t}
[\ln (\lambda \Gamma_{n}+q_n)]
=Q_{n+1}-Q_n,
\end{eqnarray}
where
\begin{eqnarray}
Q_n=V_{21}^{(m)}\Gamma_{n}
+V_{22}^{(m)}.
\end{eqnarray}
The spectral problem (3.3) can be written in the form,
\begin{equation}
\lambda\Gamma_n\Gamma_{n+1}+q_n\Gamma_{n+1}-1=0,
\end{equation}
which is a discrete Riccati equation. In order to solve the equation, we
suppose the eigenfunction $\psi_2(n,t,\lambda)$
is an analytical function of the arguments and expand $\Gamma_n$ with respect to $\lambda$ by the Taylor series
\begin{eqnarray}
\Gamma_n=\sum_{j=0}^{\infty}\lambda^{j}w_n^{(j)},
\end{eqnarray}
and then $w_n^{(j)}$ can be determined recursively as follows,
\begin{eqnarray}
w_n^{(0)}=\frac{1}{q_{n-1}},\qquad
w_n^{(j)}=\frac{-1}{q_{n-1}}\sum_{l+m=j-1}w_{n-1}^{(l)}w_n^{(m)}, \qquad j=1,2,3,.....
\end{eqnarray}
i.e.,
$$w_n^{(1)}=\frac{-1}{q_{n-2}q_{n-1}^2},\qquad
w_n^{(2)}=\frac{1}{q_{n-2}^2q_{n-1}^2}(\frac{1}{q_{n-3}}+\frac{1}{q_{n-1}}),$$
\begin{eqnarray}
w_n^{(3)}=\frac{-1}{q_{n-2}^2q_{n-1}^2}[\frac{1}{q_{n-2}q_{n-1}}(\frac{1}{q_{n-1}}+\frac{2}{q_{n-3}})+\frac{1}{q_{n-3}^2}(\frac{1}{q_{n-2}}+\frac{1}{q_{n-4}})],
\end{eqnarray}
$$................$$
Further, from equation (3.5) we have
\begin{eqnarray}
\frac{\partial}{\partial t}ln q_n+\frac{\partial}{\partial t}\sum_{k=1}^{\infty}\frac{(-1)^{k+1}\lambda^{k}}{k}\Phi^k
=Q_{n+1}-Q_n,
\end{eqnarray}
with
\begin{eqnarray}
\Phi=\sum_{j=0}^{\infty}\lambda^j\bar{w}_n^{(j)},\qquad \bar{w}_n^{(j)}=\frac{w_n^{(j)}}{q_n}.
\end{eqnarray}
Equation (3.11) leads to the form,
\begin{eqnarray}
\frac{\partial }{\partial t} \sum_{j=0}^{\infty}\lambda^{j}\alpha_n^{(j)}
=\frac{\partial }{\partial t}\alpha_n^{(0)}+ \sum_{j=1}^{\infty}(aj\lambda^{j+\beta-1}\alpha_n^{(j)}+\lambda^j\frac{\partial}{\partial t}\alpha_n^{(j)})=Q_{n+1}-Q_n,
\end{eqnarray}
where
$$\alpha_n^{(0)}=ln q_n, \qquad \alpha_n^{(1)}=\frac{1}{q_nq_{n-1}},\qquad 
\alpha_n^{(2)}=\frac{-1}{q_{n-1}^2q_n}(\frac{1}{q_{n-2}}+\frac{1}{2q_n}),$$
$$\alpha_n^{(3)}=\frac{1}{q_{n-2}^2q_{n-1}^2q_n}(\frac{1}{q_{n-1}}+\frac{1}{q_{n-3}})+\frac{1}{q_{n-2}q_{n-1}^3q_n^2}+\frac{1}{3q_{n-1}^3q_n^3},$$
\begin{eqnarray}
\alpha_n^{(j)}=\bar{w}_n^{(j-1)}-\frac{1}{2}\sum_{l_1+l_2=j-2}\bar{w}_n^{(l_1)}\bar{w}_n^{(l_2)}+\frac{1}{3}\sum_{l_1+l_2+l_3=j-3}\bar{w}_n^{(l_1)}\bar{w}_n^{(l_2)}\bar{w}_n^{(l_3)}
-.....+\nonumber\\
\frac{(-1)^{j-1}}{j-2}\sum_{l_1+l_2+...+l_{j-2}=2}\bar{w}_n^{(l_1)}\bar{w}_n^{(l_2)}....\bar{w}_n^{(l_{j-2})}+
(-1)^j(\bar{w}_n^{(0)})^{j-2}\bar{w}_n^{(1)}+\frac{(-1)^{j+1}}{j}(\bar{w}_n^{(0)})^{j}.
\end{eqnarray}
In comparison with the powers of $\lambda$ on both sides of equation (3.13),
we obtain infinitely many conservation laws for lattice hierarchy related to nonisospectral problem (1.2),
\begin{eqnarray}
\rho_{n,t}^{(i)}=J_{n+1}^{(i)}-J_{n}^{(i)}, \qquad i=0,1,2,3,......
\end{eqnarray}
{\bf 3.2} Infinitely many conservation laws of lattice hierarchies (2.3) and (2.9)\\
For lattice hierarchy (2.3), notice that
\begin{eqnarray}
Q_n=\lambda \Gamma_nEA^{(m)}(\lambda)+C^{(m)}(\lambda)
=\sum_{i=0}^{\infty}J_n^{(i)}\lambda^i,
\end{eqnarray}
where
\begin{eqnarray}
J_n^{(i)}=\left\{
\begin{array}{cc}
C_{m-i}+\sum_{s+l=i-1}w_n^{(s)}EA_{m-l},& 0\leq i \leq m+1,\\
\sum_{s+l=i-1}w_n^{(s)}EA_{m-l}, & i\geq m+2
\end{array}\right.
\end{eqnarray}
we thus obtain its infinitely many conservation laws (3.15), where the associated fluxes $J_n^{(i)}(i=0,1,2,.....)$ are presented by equation (3.17), and the conserved density
$\rho_n^{(i)}(i=0,1,2,.....)$ are written in the form,
\begin{eqnarray}
\rho_n^{(i)}=\left\{
\begin{array}{cc}
\alpha_n^{(i)}, &0\leq i \leq m+1,\\
\alpha_n^{(i)}+a(i-m-1)\int_{0}^{t}\alpha_n^{(i-m-1)}dt, & i\geq m+2,
\end{array}
\right.
\end{eqnarray}
For lattice hierarchy (2.9), notice that
\begin{eqnarray}
Q_n=\lambda \Gamma_nEA^{(m)}(\lambda)+C^{(m)}(\lambda)=\sum_{i=0}^{\infty}J_n^{(i)}\lambda^{-m+i},
\end{eqnarray}
where
\begin{eqnarray}
J_n^{(i)}=\left\{
\begin{array}{cc}
C_{i+1}+\sum_{s+l=i}w_n^{(s)}EA_{l}, & 0\leq i\leq m-1,\\
\sum_{s+l=i}w_n^{(s)}EA_{l}, & i\geq m
\end{array}
\right.
\end{eqnarray}
with $A_l=0$ for $l\geq m+1$.
Thus, lattice hierarchy (2.9) possesses infinitely many conservation laws (3.15), where the associated fluxes $J_n^{(i)}(i=0,1,2,.....)$ are described by equation (3.20), and
the conserved density $\rho_n^{(i)}(i=0,1,2,.....)$ are written in the form,
\begin{eqnarray}
\rho_n^{(i)}=\left\{
\begin{array}{cc}
(i+1)a\int_{0}^{t}\alpha_n^{(i+1)}dt, & 0\leq i\leq m-1,\\
ln q_n+(m+1)a\int_{0}^{t}\alpha_n^{(m+1)}dt,& i=m,\\
\alpha_n^{(i-m)}+(i+1)a\int_{0}^{t}\alpha_n^{(i+1)}dt, & i\geq m+1
\end{array}
\right.
\end{eqnarray}
Conserved quantities $H_i, i\geq 0$ of lattice hierarchy (2.3) possess the following forms,
\begin{eqnarray}
H_0=\sum_{n}ln q_n, \qquad H_1=\sum_{n}\frac{1}{q_nq_{n-1}},\qquad
H_2=\sum_{n}\frac{-1}{q_{n-1}^2q_n}(\frac{1}{q_{n-2}}+\frac{1}{2q_n}),\nonumber\\
H_3=\sum_{n}\frac{1}{q_{n-2}^2q_{n-1}^2q_n}(\frac{1}{q_{n-1}}+\frac{1}{q_{n-3}})+\frac{1}{q_{n-2}q_{n-1}^3q_n^2}+\frac{1}{3q_{n-1}^3q_n^3},
\end{eqnarray}
$$........................$$
$$H_{m+1}=\sum_{n}\alpha_n^{(m+1)}, \qquad H_i=\sum_{n}[\alpha_n^{(i)}+a(i-m-1)\int_{0}^{t}\alpha_n^{(i-m-1)}dt], i\geq m+2.$$
For lattice hierarchy (2.9), conserved quantities $H_i, i\geq 0$ can be described by
\begin{eqnarray}
H_i=\left\{
\begin{array}{cc}
\sum_{n}(i+1)a\int_{0}^{t}\alpha_n^{(i+1)}dt, & 0\leq i\leq m-1,\\
\sum_{n}[ln q_n+(m+1)a\int_{0}^{t}\alpha_n^{(m+1)}dt],& i=m,\\
\sum_{n}[\alpha_n^{(i-m)}+(i+1)a\int_{0}^{t}\alpha_n^{(i+1)}dt], & i\geq m+1
\end{array}
\right.
\end{eqnarray}
{\bf Example 1} For lattice equation (2.6), 
since
\begin{eqnarray}
Q_n=-\lambda \Gamma_n+(\frac{n}{2}+\frac{1}{4})a-c_1(-1)^{n}-\sum_{k=0}^{\infty}(-1)^{k}q_{n+k}
\end{eqnarray}
it possesses infinitely many conservation laws (3.15), where 
the associated fluxes $J_n^{(i)}, i\geq 0$ are written by the
following equations, respectively,
$${J}_n^{(0)}=(\frac{n}{2}+\frac{1}{4})a-c_1(-1)^{n}-\sum_{k=0}^{\infty}(-1)^{k}q_{n+k},$$
$${J}_n^{(1)}=\frac{-1}{q_{n-1}}, \qquad 
{J}_n^{(2)}=\frac{1}{q_{n-2}q_{n-1}^2},\qquad
{J}_n^{(3)}=\frac{-1}{q_{n-2}^2q_{n-1}^2}(\frac{1}{q_{n-3}}+\frac{1}{q_{n-1}}),$$
\begin{eqnarray}
{J}_n^{(4)}=\frac{1}{q_{n-2}^2q_{n-1}^2}[\frac{1}{q_{n-2}q_{n-1}}(\frac{1}{q_{n-1}}+\frac{2}{q_{n-3}})+\frac{1}{q_{n-3}^2}(\frac{1}{q_{n-2}}+\frac{1}{q_{n-4}})],
\end{eqnarray}
$${J}_n^{(i)}=-w_n^{(i-1)}, \qquad i\geq 5$$
For lattice equation (2.7),
notice that
\begin{eqnarray}
Q_n=(\lambda^2EA_{-1}+\lambda EA_0)\Gamma_n+\lambda C_{-1}-B_0,
\end{eqnarray}
hence it admits infinitely many conservation laws (3.15), where the
associated fluxes $J_n^{(i)}, i\geq 0$ are given by,
$$ J_n^{(0)}=-B_0,\qquad
J_n^{(1)}=\frac{EA_0}{q_{n-1}}+C_{-1}, \qquad
J_n^{(2)}=\frac{EA_{-1}}{q_{n-1}}-\frac{EA_0}{q_{n-2}q_{n-1}^2},$$
\begin{eqnarray}
J_n^{(i)}=w_n^{(i-2)}EA_{-1}+w_n^{(i-1)}EA_{0}, \qquad i\geq 3
\end{eqnarray}
{\bf Example 2} For lattice equation (2.11), 
note that
\begin{eqnarray}
Q_n=\frac{1-(n+1)a}{q_n}\Gamma_n,
\end{eqnarray}
we thus obtain its infinitely many conservation laws (3.15), where the associated fluxes $J_n^{(i)},i\geq 0$ have the formula,
\begin{eqnarray}
J_n^{(i)}=\frac{1-(n+1)a}{q_n}w_n^{(i)}, \qquad i\geq 0
\end{eqnarray}
For lattice equation (2.12), we have
\begin{eqnarray}
Q_n=(EA_0\lambda^{-1}+EA_1)\Gamma_n-B_1\lambda^{-1},
\end{eqnarray}
so, its infinitely many conservation laws (3.15) is given, where 
the associated fluxes $J_n^{(i)},i\geq 0$ are described by
\begin{eqnarray*}
J_n^{(0)}=-a\sum_{k=-1}^{\infty}\frac{1}{q_{n+k}q_{n+k+1}}, 
\end{eqnarray*}
\begin{eqnarray*}
J_n^{(1)}=\frac{na-1}{q_{n-1}q_n^2}(\frac{1}{q_{n+1}}+\frac{1}{q_{n-1}})-\frac{1-(n+1)a}{q_{n-2}q_{n-1}^2q_n}-\frac{2a}{q_{n-1}q_n}\sum_{k=1}^{\infty}\frac{1}{q_{n+k}q_{n+k+1}}-\frac{1-(n+1)a}{q_{n-2}q_{n-1}^2q_n}
\end{eqnarray*}
\begin{eqnarray}
J_n^{(i)}=\sum_{s+l=i}w_n^{(s)}EA_{l}, \qquad i\geq 2,
\end{eqnarray}
here $A_l=0$ for $l\geq 2$.
\section{Conclusions}
It is well known that the Lax pairs and infinitely many conservation laws are two important integrable properties for a discrete lattice system. Specially, infinitely many conservation laws for the 
lattice hierarchy with n-dependent coefficient has little work in the literature.  
In this article, by means of discrete zero curvature representation and constructing opportune time evolution equations, two new discrete integrable lattice hierarchies with n-dependent coefficients are proposed, which associated with a new discrete Schr\"{o}dinger nonisospectral problem. Further,
it has been shown that lattice hierarchy (2.9) is equivalent to the positive Volterra hierarchy with n-dependent coefficients and lattice hierarchy (2.3) related to isospectral problem is equivalent to the negative Volterra hierarchy.
We also demonstrate the existence of infinitely many conservation laws for the proposed two lattice hierarchies and give the corresponding conserved densities and the associated fluxes formulaically.
Thus their integrability is confirmed. The meaning of lattice hierarchy (2.3) related to nonisospectral problem is worth further investigation.
\vskip 2mm
\noindent
{\bf {Acknowledgements}}\\
This work was supported by Research Grant Council of Hong Kong UGC, HKBU2047/02P.

{\small
\begin{thebibliography}{s2}
\bibitem{s1}M. Boiti, M. Bruschi, F. Pempinelli and B. Prinari, J. Phys. A: Math. Gen.
{\bf 36} (2003) 139-149.
\bibitem{s2}M. Bruschi and O. Ragnisco, J. Phys. A: Math. Gen. {\bf 14}( 1981) 1075.
\bibitem{s3}M. Bruschi and O. Ragnisco, Lettere al Nuovo Cimento {\bf 29} (1980) 321.
\bibitem{s4}A. Shabat, in Nonlinearity, Integrability and all That. Twenty Years After NEEDS'79, ed. M. Boiti, L. Martina, F. Pempinelli, B. Prinari and G. Soliani (Singapore: World Scientific, 2000) 331.
\bibitem{s5}M. Boiti, F. Pempinelli, B. Prinari and A. Spire, Inverse Problems {\bf 17} (2001) 515.
\bibitem{s6}Z.N Zhu, H. W Tam and Q Ding, Phys. Lett. A, {\bf 300} (2003)
\bibitem{s7}T. Tsuchida, H. Ujino and M. Wadati: J. Math. Phys. {\bf 39} (1998) 4785.
\bibitem{s8}T. Tsuchida, H. Ujino and M. Wadati: J. Phys. A {\bf 32} (1999) 2239.
\bibitem{s9}Z.N Zhu, Xiaonan Wu, Weimin Xue and Z.M Zhu: Phys. Lett. A: {\bf 296}(2002) 280. 
\bibitem{s10}Z.N Zhu, Weimin Xue, Xiaonan Wu and Z.M Zhu: J. Phys. A: Math. Gen. {\bf 35}(2002) 5079. 
\bibitem{s11}Blaszak M and Marciniak K: J. Math. Phys. {\bf 35} (1994) 4661.
\bibitem{s12}D. Levi and A. M. Grundland, J. Phys.A: Math. Gen. {\bf 35}(2002) L67.
\bibitem{s13}G. M. Pritula and V. E. Vekslerchik, J. Phys.A: Math. Gen. {\bf 36}(2003) 213.
\end{thebibliography}
}
\end{document}